\documentclass[aps,prl,twocolumn,letterpaper,superscriptaddress]{revtex4}

\usepackage{epsf,rotate,color}
\usepackage{amsmath}
\usepackage{amssymb}
\usepackage{latexsym}
\usepackage{pstricks}
\usepackage{graphics}
\usepackage{lscape}
\usepackage{rotating}
\usepackage[normalem]{ulem}

\begin{document}
\title{Adaptive resolution molecular dynamics simulation through coupling to an internal particle reservoir}
\author{Sebastian Fritsch}
\thanks{Shared First Authorship: S.~Fritsch \& S.~Poblete}
\affiliation{Max Planck Institut f{\"u}r Polymerforschung,
Ackermannweg 10, D-55128 Mainz, Germany}
\author{Simon Poblete}
\thanks{Shared First Authorship: S.~Fritsch \& S.~Poblete}
\affiliation{Max Planck Institut f{\"u}r Polymerforschung,
Ackermannweg 10, D-55128 Mainz, Germany}
\author{Christoph Junghans}
\affiliation{Max Planck Institut f{\"u}r Polymerforschung,
Ackermannweg 10, D-55128 Mainz, Germany}
\author{Giovanni Ciccotti}
\affiliation{EMPS Building, Room 302/B UCD, School of Physics, University College Dublin, Belfield, Dublin 4, Ireland and
Dipartimento di Fisica and CNISM, Universit{\'a} ``La Sapienza'', Piazzale Aldo Moro 2, 00185 Rome, Italy}
\author{Luigi Delle Site}
\email{dellsite@mpip-mainz.mpg.de}
\affiliation{Max Planck Institut f{\"u}r Polymerforschung,
Ackermannweg 10, D-55128 Mainz, Germany}
\author{Kurt Kremer}
\email{kremer@mpip-mainz.mpg.de}
\affiliation{Max Planck Institut f{\"u}r Polymerforschung,
Ackermannweg 10, D-55128 Mainz, Germany}

\begin{abstract}
For simulation studies of (macro) molecular liquids it would be of significant interest to be able to adjust or increase the level of resolution within one region of space, while allowing for the free exchange of molecules between open regions of different resolution or representation. We generalize the adaptive resolution idea and suggest an interpretation in terms of an effective generalized grand canonical approach. The method is applied to liquid water at ambient conditions.
PACS numbers: 02.70.Ns,  61.20.Ja, 80.82,  05.10.-a, 05.20.Gg
\end{abstract}

\maketitle
Many (macro-) molecular systems display phenomena and properties that are inherently multiscale in nature, and thus particularly challenging for both theoretical and experimental investigations. Here computer simulations represent a powerful tool of investigation, though truly large all-atom simulations of complex molecular systems are neither possible, because of CPU requirements, nor in many cases very useful, as they often produce an excess of data.
To the aim of linking large scale generic properties and local specific chemistry, a variety of scale bridging simulation techniques, ranging from hierarchical parameterizations of models covering different levels of resolution, see e.g. \cite{prllu,jacs1,jacs2,
karsten1,karsten2,florianjcp,vothbook}, to interfaced layers of different resolutions (see e.g.
\cite{kax,Rottler:2002,Csanyi:2004,Laio:2002,Jiang:2004,Lu:2005,  villa:2004, neri:2005,delgadobuscalioni:2003, koumoutsakos:2005, lyman:2006}) have been developed.

As most techniques are sequential -- the whole system is treated on one level of resolution at a time -- switching between resolutions
may be performed only for the whole system. In many cases however it would be
desirable to adjust the level of resolution on the fly only in a~smaller well defined region of space, i.e. considering more details, according to the problem of interest, while keeping the larger surrounding on a coarser, computationally much more efficient level. This idea has been successfully employed for problems of crack propagation in hard matter~\cite{rudd}. However for soft matter or liquids inherent molecular number fluctuations pose special challenges. They require a simulation setup, which allows molecules or parts of large molecules to cross boundaries between areas of different representation without any barrier, keeping the overall thermodynamic equilibrium intact. While crossing between regions of different resolution, they either become more coarse-grained or resume more atomistic detail that is they acquire or lose degrees of freedom (DOFs). In this context, adaptive resolution MD simulations would represent a natural Grand Canonical set up: An atomistic region/cavity is coupled to a large, ideally infinite coarse-grained region acting as a particle/molecule reservoir. In general for such a coupling scheme the only requirement is that the coarse grained surrounding supplies/removes molecules and heat in a way that in the atomistic region  $\mu$,V and T remain constant.
The adaptive resolution method (AdResS, Adaptive Resolution Scheme), which allows one to  couple two systems with different resolution within one MD simulation~\cite{jcp,pre1,annurev,simonjcp,adresso}, describes a first step in this direction. Molecules change their resolution and thus their number of DOFs when moving through space and passing from one region through a transition zone into the other region. The method has successfully been extended into the quantum \cite{adolfoprl} and the continuum \cite{delgado:2008,delgado:2009} region. A different approach based on the interpolation of Lagrangians can be found in Ref.~\cite{heyden:2008}
In the following we will generalize the AdResS concept and show that the atomistic/ fine grained subsystems can be viewed as a system in contact with a (huge) coarse-grained particle/molecule reservoir, thus effectively representing a $\mu$VT ensemble on the level of the high resolution subsystem, while the overall systems remains to be an NVT ensemble. Actually,
by linking the coarse grained region in addition to a continuum, as shown in Refs.~\cite{delgado:2008,delgado:2009}, this is easily extended to
a $\mu$VT ensemble on the level of the \emph{whole} system. We now first shortly review the adaptive resolution method AdResS and the concept of thermodynamic force which assures overall thermodynamic equilibrium. We then extend this to a more general coupling scheme and apply it to the example of an adaptive resolution simulation of liquid water, where the coarse-grained water model matches the compressibility of the all-atom model, while the pressure is higher by three orders of magnitude.

\begin{figure}
\includegraphics[width=\columnwidth]{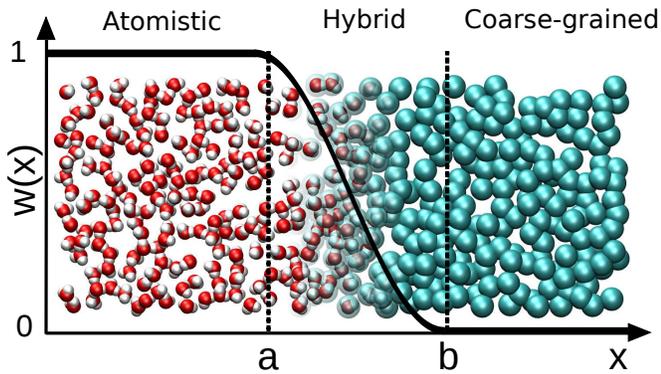}
\caption{Pictorial representation of the adaptive simulation box and
local molecular representation. On the left, is the low resolution (coarse-grained) region, indicated by $B$, and on the right, is the high resolution (atomistic) region {A}. In the middle is the transition (hybrid) region $H$ with the switching function $w(x)$ (curve in grey), \label{box}}
\end{figure}

In the AdResS method molecules smoothly change their level of
resolution as a function of position by moving through a transition region as illustrated in Fig.\ref{box}.
In order to accomplish the process described above, two different representations are interpolated by a transition function $w(x)$ with $w=1$ in region $A$, the fine grained more microscopic or all-atom one, and $w=0$ in the coarse-grained region $B$, respectively. Further let us assume that a particle in $B$ is the coarse-grained representation of a collection of atoms in region $A$ and the position of the coarse particle is the center of mass of the corresponding atomistic molecule.
With this in mind two molecules $\alpha, \beta$ in region $A$ (fine-grained, $w(X_\alpha)=w(X_\beta)=1$, $X_\alpha$, $X_\beta$ being the x-coordinates of the centers of mass of the two molecules, cf Fig.~\ref{box}) interact via the standard forces between all individual atoms of the molecules;
in region $B$ (coarse-grained, $w(X_\alpha)=w(X_\beta)=0$) two coarser molecules interact with each other via central forces; just as in a bulk system of all atom molecules or coarse grained molecules, respectively.
In all other cases of \textbf{mixed resolution} interactions, these two forces are interpolated as illustrated in Fig.~\ref{box} so that the total force
$\mathcal{F}_{\alpha\beta}$ between two different molecules  consist of two parts. The all-atom contribution reads:
\begin{equation}
{\bf \mathcal{F}}^A_{\alpha\beta}=w(X_\alpha)w(X_\beta)\sum_{i,j}{\bf F}^A_{\alpha_i\beta_j}~~~~~\alpha\neq\beta
\label{forces2}
\end{equation}
While the coarse-grained parts is:
\begin{equation}
{\bf \mathcal{F}}^B_{\alpha\beta}=\left(1-w(X_\alpha)w(X_\beta)\right){\bf F}^B_{\alpha\beta}~~~~~\alpha\neq\beta
\label{forces1}
\end{equation}
${\bf F}^A_{\alpha_i\beta_j}$ is the force between atom $i$ of molecule $\alpha$ and atom $j$ of molecule $\beta$ and ${\bf F}^B_{\alpha\beta}$ is the coarse-grained center of mass -- center of mass interaction. Note that in the transition region the intra-molecular atom-atom interactions are left unaffected. For technical details of the implementation we refer to the supporting material.
Altogether we simulate a system at constant volume, partitioned in two
different (ideally macroscopic) regions (coarse-grained and fine-grained) by a transition region of finite width. This region facilitates the exchange of molecules and ensures equilibrium between the coarse and fine grained regions, moreover, by construction (as explained below), should not affect the density fluctuations (compressibility).
Beyond the crucial technical task of assuring the thermodynamically correct coupling between the two large regions, the transition region does not have, a priori, any thermodynamic/physical meaning. An important characteristic of the adaptive approach is that while the overall number of molecules is conserved, this is not the case for the number of atoms in the individual regions.
Eventually we aim for a flat density profile throughout the whole simulation box, including the transition region, so that we have a
thermodynamic state point well defined without (kinetic) barriers for the exchange of molecules.

In the same spirit, the temperature has to be well defined and equal throughout the system. However, while this is simple to achieve in the "bulk" regions, the transition region requires some care since in this region a change in the number of degrees of freedom occurs.
By switching from coarse-grained to atomistic or vice versa, degrees of freedom (DOFs) are reintroduced or removed and the related equilibration is accomplished by a  "local" thermostating procedure (for details see the supporting material). The approach presented above, paves the way to an open system simulation scheme within a generalized Grand Canonical framework. 

A common method to derive coarse grained interactions is based on matching the center of mass radial distribution functions between the molecules in the two systems. As a result the compressibilities $\kappa$ remain unchanged~\cite{hansentheory} by construction while the pressures in the fine and coarse-grained representation can be different~\cite{fm5}. In terms of the grand potential $-pV$ this reads (for identical volumes)
\begin{equation}
p_{A}(\mu_A, T)V \neq p_{B}(\mu_B, T)V;~~~~~
\kappa_{A} =  \kappa_{B},
\label{kappa}
\end{equation}
where $\mu_A$ and $\mu_B$ are the respective chemical potentials. This will result in density variations in the system. Instead we would like to have a constant density $\rho_0$ throughout the whole simulation box, while keeping identical compressibilities. Different pressures at the boundaries of the transition region will produce a drift force on the molecules, leading to an unphysical inhomogeneous system. Different compressibilities would affect the molecule number fluctuations in the - still finite - regions.  Thus we keep $\kappa_{A} =  \kappa_{B}$ and propose to exactly compensate this pressure generated drift force by a thermodynamic force ${\bf F}_\text{th}(x)$ acting on the molecules
and Eq. \ref{kappa} can then be rewritten as (see Fig.~\ref{box})
\begin{equation}
\left(p_{A}(\mu_A, T) + \frac{\rho_{0}}{M_\alpha}\int_a^b  {\bf F}_\text{th}(x) \text{d}x\right) V =p_{B}(\mu_B, T)V
\label{kappa2}
\end{equation}
with $\rho_{0}$ being the reference, uniform, molecular density.
Eq.\ref{kappa2} tells that the subsystem $A$ ($B$) is in equilibrium with a reservoir $\mu_B$ ($\mu_A$) respectively, notwithstanding the fact that
the pressures, $p_{A}$ and $p_{B}$, and local chemical potentials, $\mu_A$ and $\mu_B$, may be different. To overcome these variations, the related thermodynamic work, which exactly compensates any related drift force between the different regions, is provided/absorbed by ${\bf F}_\text{th}$.
Let the all-atom system, at a given density $\rho_0$, serve as reference system, then the force on a~molecule with mass $M_\alpha$ balancing $-\nabla p(x)$, reads
\begin{equation}
{\bf F}_\text{th}(x) = \frac{M_\alpha}{\rho_0} \nabla p(x)
\label{thermo_force}
\end{equation}
where $p(x)$ is the local pressure at the target density $\rho_0$. ${\bf F}_\text{th}$ is applied for $a<x<b$, where $a$ ($b$) is the left (right) boundary of the transition region.
In Eq.\ref{thermo_force}, $p(x=a)=p_{A}$ and $p(x=b)=p_{B}$.
Now the density will remain unchanged. With this generalization the original equation for the coarse-grained forces in the transition region, Eq.~\ref{forces1} is changed so that the force ${\bf \mathcal{\hat{F}}}^B_{\alpha}$ acting on the center of mass of molecule $\alpha$ reads:
\begin{equation}
{\bf \mathcal{\hat{F}}}^B_{\alpha}=\sum_{\beta}{\bf \mathcal{F}}^B_{\alpha\beta} + {\bf F}_\text{th}(X_\alpha)
  \label{AdResS-gen}
\end{equation}
The above line of arguments does not require the
pressures in the two (''macroscopic") regions, $p_{A}$ and $p_{B}$, to be the
same, allowing to couple rather different systems. Formally this simply means
that the thermodynamic force as well as the thermostat can perform work on the molecules to adjust the virial pressure and the thermal energy, respectively, while passing from one region to the other.
This procedure allows one to  couple almost arbitrary systems and keep them in equilibrium with respect to each other. For typical coarse-graining procedures like (iterative) Boltzmann inversion or Reverse Monte Carlo which are mostly based on radial distribution functions and reproduce structural rather than thermodynamic aspects very well, this is of special advantage.\\
Unfortunately the optimized pressure profile $p(x)$ is not directly accessible as it has to be measured under the constraint of enforcing the equilibrium density $\rho_0$ in the system. Thus we developed an iterative approach based on the density profile. Within a linear approximation a first estimate of the local pressure $p(x,\rho(x))$  for an enforced overall constant density $\rho_0$ is:
\begin{equation}
p(x,\rho(x))) \approx p_{A} + \frac{1}{\rho_0 \kappa_T}\left(\rho_0-\rho(x)\right)
\label{linear_expansion}
\end{equation}
where $\kappa_T$ is the (constant) isothermal compressibility, and $\rho(x)$ is the, non uniform, density to which the system would adjust itself to without applying an external force.
In order to enforce the uniform density $\rho_0$, Eq. \ref{linear_expansion} and Eq. \ref{thermo_force} allow to obtain an initial thermodynamic force, that can be refined iteratively through
\begin{equation}
{\bf F}_\text{th}^{i+1}({x}) = {\bf F}_\text{th}^{i}({x}) - \frac{M_\alpha}{\rho_0^2 \kappa_T} \nabla \rho^{i}({x}),
\label{iterative_thermo_force}
\end{equation}
until the system evolves to the target uniform density.
Here for $i=1$, we have $\rho^{1}({x})=\rho({x})$ and ${\bf F}_\text{th}^{1}({x})=0$.
The prefactor $1/{\rho_0^2\kappa_T}$ can be interpreted as a variation of local chemical potential, by the identity \cite{hansentheory}
\begin{equation}
\left({\frac{\partial\mu}{\partial\rho}}\right)_{V,T}=\frac{1}{\rho^{2}\kappa_{T}},
\label{eq:dmu_drho}
\end{equation}
a fact explored already before in this context \cite{simonjcp}.
In the following we will demonstrate the power of this concept for the illustrative case of liquid water, where the compressibility of the SPC/E water and a structure based coarse-grained model are the same, while the pressure in the coarse-grained system is 6000\;bar\cite{wang2009comparative}.

\begin{figure}
\begin{center}
\includegraphics{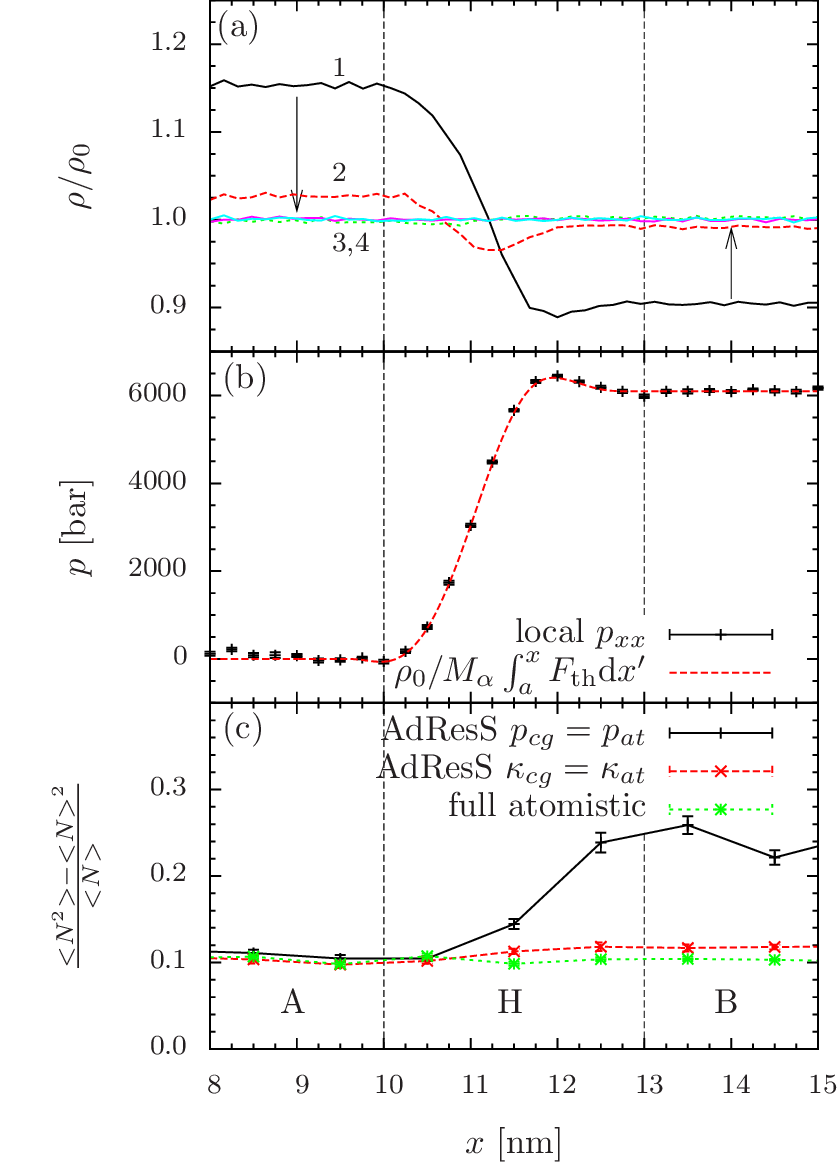}
\caption{(a) Density profile for the water simulation for the uncorrected AdResS simulation of water and several iterations of the thermodynamic force with $A$ being the all-atom SPC/E region and $B$ the coarse-grained region. The final density profile shows deviation of less than 0.4 \%. (The arrow indicates the development of the density with the iteration.)
(b) Compensation of the local pressure change between regions by the thermodynamic force.
(c) Molecule number fluctuations in slabs (width $\Delta x=1\text{nm}$) along the x-axis for pressure corrected and uncorrected coarse-grained models.}
\label{water_density_and_partfluc}
\end{center}
\end{figure}

Coarse graining liquid water with a focus on the liquid structure usually leads to models for which the pressure is significantly higher while the compressibility remains essentially unchanged. Thus pressure correction terms are often used, which add linear terms to the coarse-grained potential. The compressibility is then affected as $\kappa_T \propto \int r^2 (g(r)-1) \text{d}r$ and small deviations at high $r$ values can have dramatic effects ~\cite{wang2009comparative}. However, for adaptive resolution methods, but also solvent-solute systems, it is important that the compressibility in the coarse-grained and all-atom regions is the same.  This minimizes finite size effects for our (in most cases on purpose as small as possible) atomistic region, imposed by the - still finite - CG region.
For many applications in computational biology, but also for studies of water itself, the SPC/E model is employed \cite{berendsen1987missing}.
A standard structure based coarse-grained model perfectly fits the radial
distribution functions, but does not comply with the tetrahedral packing
\cite{wang2009comparative, watertetra2}. It has -within the
error bars- the same compressibility but a very high pressure of
about 6000\;bar, making it an ideal test case for the thermodynamic force principle described above. If not properly accounted for, this pressure difference between the different models would lead to density distortions throughout the simulation box and create a transient flow of molecules from the coarse-grained to the atomistic region. Using
$\kappa_T^{A}$ at the outset of the iteration scheme of
Eq. \ref{iterative_thermo_force} as described before, after only four
iterations for the thermodynamic force we arrive at a flat density profile, as illustrated in
Fig. \ref{water_density_and_partfluc}.

\begin{figure}
\begin{center}
\includegraphics{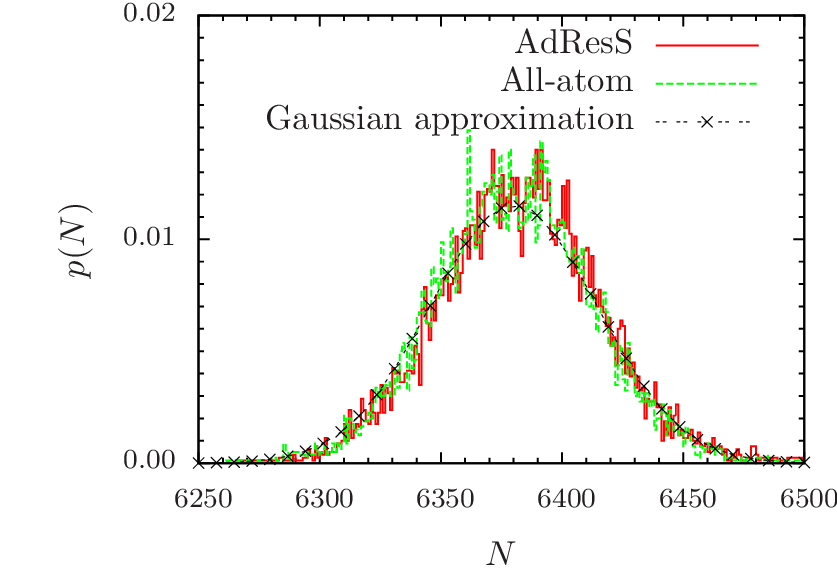}
\caption{Particle number distribution in a slab of width $l=4\;\text{nm}$ and for an atomistic zone in adaptive resolution with the same width. Both distributions match the expected Gaussian behavior with mean $\mu=\langle N \rangle$ and variance $\sigma^2=\rho k_{B}T\kappa_{T} \langle N \rangle$
}
\label{distribution}
\end{center}
\end{figure}

The thermodynamic force creates a stationary state with the correct target density, removing both the effect of the pressure difference of the models on the distribution of molecules in either region and the artifacts from the change of resolution in the transition region.
The advantages are also clearly displayed, when looking at local density fluctuations. While in the thermodynamic limit the molecule number fluctuations are related to the compressibility through
$
\rho k_B T\kappa_{T}=\frac{\langle N^{2}\rangle-\langle N \rangle^{2}}{\langle N \rangle},
$
we here analyze an effective compressibility in terms of molecule number fluctuations in slabs along the axis of resolution change.
The increased compressibility for a pressure corrected coarse-grained SPC/E water model leads to an increase in the  local density fluctuations.
Within our setup we do not alter the effective compressibility anywhere in the system as shown in Fig. \ref{water_density_and_partfluc}c. Actually the molecule number fluctuations in subvolumes of region $A$, are the same as in similar subvolumes of a full all-atom simulation.
Hence the particle distribution in region $A$ shows the expected Gaussian distribution as it is the case in the full all-atom simulation, shown in Fig. \ref{distribution}.
Leveling out any pressure undulations in the transition region, while keeping the compressibilities the same also means that there is no barrier for molecule exchange between regions (see supporting information).
Coupling two systems with different pressures means that the thermodynamic force performs work on the molecules, when they cross from one region to the other. In practice the integral over the thermodynamic force over the transition region does not have to be zero, which would be the case if the pressures on either side were the same. Thus for the present setup the thermodynamic force has to exactly compensate the local pressure variation due to the constant density. The comparison of the final local pressure (see supporting information for details) and the thermodynamic force as derived in form of an external local pressure field is shown in Fig. \ref{water_density_and_partfluc}b. They perfectly match, illustrating the overall consistency of the approach. Thus by applying the thermodynamic force we arrive at a robust and efficient simulation setup, which enables us within a single MD simulation to keep rather different systems with quite different intrinsic pressures but identical compressibilities in equilibrium with each  other.


The fact that the number of molecules at a given resolution fluctuates suggests to view each subsystem as a reservoir of molecules for the other. Strictly speaking, to interpret such a setup as a Grand Canonical ensemble would require macroscopic reservoirs. In our case it is obvious that the number of molecules, and thus the reservoir, is finite.  In fact, we have shown that the average and the fluctuating number of molecules in each subsystem and even in smaller subregions throughout the whole simulation box are the same as analyzed in a full atomistic system.
For the general case where the all-atom as well as the coarse-grained system are at the same (or even well defined different) densities the simulation setup leads to a situation, where a difference in the excess chemical potential $\mu$ (the temperature part for the internal degrees of freedom is taken care of by the thermostat) between the regions, is compensated by the work performed by the thermodynamic force. Thus on the level of the (large) subsystems we run a ($\mu,V,T$) MD simulation, while we have a ($N,V,T$) ensemble globally.
The extreme example of liquid SPC/E water at temperature 300K and pressure 1\;bar, where the coarse-grained model at the same temperature and density has a pressure of more than $\approx$ 6000\;bar, shows the power of the present approach.
In addition, as mentioned before \cite{delgado:2008,delgado:2009} one can link the coarse-grained region quite easily to a continuum domain, making it globally Grand Canonical.
This opens completely new pathways to study systems, where a rather high level of detail is only required within a rather restricted area, while at the same time the exchange with the surrounding is essential. First examples of applications are (small) organic molecules such as amino acids or peptides in mixed solvents, e.g. water and urea. For peptides the conformation is expected to strongly couple to the urea density close to the peptide. There the present setup facilitates very fast and efficient high precision calculations of Kirkwood Buff integrals to estimate solvation free energies. More generally, among many others, potential applications include growth and structure formation phenomena in soft matter where locally the assembly process needs details of the molecular structure (templated crystal growth, bio-mineralization etc.) and/or a controlled flux of additional molecules is provided.
In conclusion we have derived a thermodynamically sound, technically
robust, and easy to implement way to have, in a single MD simulation one (small) system coupled with a different (large) system in Grand Canonical equilibrium. We exemplified this for the case of an atomistic system in Grand Canonical equilibrium with a source of atomistic molecules provided by a bath of coarse grained molecules. This is possible due to the work provided by the thermodynamic force in the transition zone, acting like an active membrane, facilitating the transformation of particles/molecules.

\emph{Acknowledgment}
We thank B. D\"unweg and M. Praprotnik for valuable discussions. We thank M. Deserno and D. Donadio for comments on the manuscript. Part of the work has been performed with the Multiscale Materials Modeling Initiative of the Max Planck Society. S. P. thanks the DAAD-Conicyt grant for financial support. G.C. wish to acknowledge financial support from SFI Grant 08-IN.1-I1869 and from the Istituto Italiano di Tecnologia (IIT) under the seed project grant 259, SIMBEDD. C.J. was financially supported by SFB 625.

\end{document}


\title{Supporting Material: Adaptive resolution molecular dynamics simulation through coupling to an internal particle reservoir}
\author{Sebastian Fritsch}
\thanks{Shared First Authorship: S.~Fritsch \& S.~Poblete}
\affiliation{Max Planck Institut f{\"u}r Polymerforschung,
Ackermannweg 10, D-55128 Mainz, Germany}
\author{Simon Poblete}
\thanks{Shared First Authorship: S.~Fritsch \& S.~Poblete}
\affiliation{Max Planck Institut f{\"u}r Polymerforschung,
Ackermannweg 10, D-55128 Mainz, Germany}
\author{Christoph Junghans}
\affiliation{Max Planck Institut f{\"u}r Polymerforschung,
Ackermannweg 10, D-55128 Mainz, Germany}
\author{Giovanni Ciccotti}
\affiliation{EMPS Building, Room 302/B UCD, School of Physics, University College Dublin, Belfield, Dublin 4, Ireland and
Dipartimento di Fisica and CNISM, Universit{\'a} ``La Sapienza'', Piazzale Aldo Moro 2, 00185 Rome, Italy}
\author{Luigi Delle Site}
\email{dellsite@mpip-mainz.mpg.de}
\affiliation{Max Planck Institut f{\"u}r Polymerforschung,
Ackermannweg 10, D-55128 Mainz, Germany}
\author{Kurt Kremer}
\email{kremer@mpip-mainz.mpg.de}
\affiliation{Max Planck Institut f{\"u}r Polymerforschung,
Ackermannweg 10, D-55128 Mainz, Germany}

\maketitle
\section{Force calculation}
In Eqn.~1 and Eqn.~2 of the original paper the interactions between molecules $\alpha$ and $\beta$ subject to the interpolation of forces is written in a rather general form as
\begin{equation}
{\bf \mathcal{F}}^B_{\alpha\beta}=\left(1-w(X_\alpha)w(X_\beta)\right){\bf F}^B_{\alpha\beta}
\label{forces1}
\end{equation}
for the center of mass force and
\begin{equation}
{\bf \mathcal{F}}^A_{\alpha\beta}=w(X_\alpha)w(X_\beta)\sum_{i,j}{\bf F}^A_{\alpha_i\beta_j} ~~~~~\alpha\neq\beta
\label{forces2}
\end{equation}
for the atom level interactions. From there one arrives at the interpolation formula for the total force between two molecules used in previous publications:
\begin{equation}
{\bf F}_{\alpha\beta}=w(X_\alpha)w(X_\beta){\bf F}_{\alpha\beta}^A+(1-w(X_\alpha)w(X_\beta)){\bf F}_{\alpha \beta}^B
\end{equation}
From this it is possible to see that, (i) the force ${\bf F}^A_{\alpha\beta}$ in the hybrid region is actually a function of all the atomic coordinates $\{r_i\}_{i\in\alpha}$ and $\{r_j\}_{j\in\beta}$
and hence need not to be in the direction of the connection vector $R_{\alpha\beta}$ of the centers of mass and can lead to e.g. rotation of the molecule.
(ii) Since the intra molecular forces do not contribute to the total force on a molecule, we do not have to include the intramolecular interactions in the interpolation scheme; they are kept acting on the individual atoms throughout the atomistic and transition regions just as they are.\\

In general Eqns.~1,2 describe four different cases for the forces between two different molecules, which apply to the whole simulation box:\\
(i) For $w(X_\alpha)=w(X_\beta)=0$ two coarse-grained molecules interact via the effective coarse-grained interaction
\begin{equation}
{\bf \mathcal{F}}_{\alpha\beta}={\bf F}^B_{\alpha\beta},
\label{f1}
\end{equation}
leading only to a translational motion of the molecules which are now reduced to point particles. In this regime there is no intermolecular or intramolecular atomistic contribution.\\
(ii) For $w(X_\alpha)=w(X_\beta)=1$ both molecules are in the atomistic region. Thus atom $i$ of molecule $\alpha$ interacts with all atoms $j$ of molecule $\beta$
\begin{equation}
{\bf \mathcal{F}}_{\alpha_i,\beta}= \sum_{j\in\beta}{\bf F}^A_{\alpha_i\beta_j}.
\label{f2}
\end{equation}
The situation is that of a standard all-atom simulation with forces acting on each atom.
\\
(iii) For $0 < w(X_\alpha) < 1$ and $w(X_\beta)\neq 0$ at least molecule $\alpha$ is in the transition regime and $\beta$ is not in the coarse-grained region. Then both parts of the interpolation contribute to the motion of the molecule, hence the coarse-grained part is also contributing to the force between the atoms. Following Eqns.~1,2 the force on atom $\alpha_i$ originating from all atoms of molecule $\beta$ reads:
\begin{equation}
{\bf \mathcal{F}}_{\alpha_i,\beta}= w(X_\alpha)w(X_\beta)\sum_{j\in\beta}{\bf F}^A_{\alpha_i\beta_j}
+ (1-w(X_\alpha)w(X_\beta))\frac{m_i}{M_\alpha}{\bf F}^\text{B}_{\alpha\beta}.
\label{f3}
\end{equation}
The first part describes the standard atom atom interactions, while the second contains the distribution of the coarse grained interaction onto the individual atoms. By this the coarse grained interactions are naturally incorporated into the all atom resolution, which we keep throughout the whole transition region. \\
(iv) For $w(X_\alpha) \neq 0$ and $w(X_\beta)= 0$ only the second part of Eqn.~\ref{f3}, the center of mass contribution, survives and has to be distributed to the atoms of molecule $\alpha$, which gives
\begin{equation}
{\bf \mathcal{F}}_{\alpha_i,\beta}= \frac{m_i}{M_\alpha}{\bf F}^\text{B}_{\alpha\beta}
\label{f4}
\end{equation}
for atom $\alpha_i$.\\
With this scheme all forces between atoms and molecules are well defined throughout the whole simulation box.\\
As mentioned in the main text, the thermostat needs some attention.
Kinetic energy and the related temperature are both well defined in the coarse-grained and the atomistic region. More attention is required in the transition region. By switching from coarse-grained to atomistic or vice versa, degrees of freedom (DOFs) (including their related thermal energies) are introduced or lost.
The related contribution to the thermal energy will be furnished by a~standard local thermostat as has been discussed in Refs.~\cite{annurev,pre2,jpa,adresso}.
Thus the molecules/particles are ''handed over" properly thermalized at the boundaries of the transition region. In summary, one can apply any standard local thermostats like, e.g, a Langevin thermostat, which automatically take care of the ''latent heat"\cite{pre2,jpa} related to the fading DOFs. AdResS simulations from the point of view of the whole system correspond to simulations of a $NVT$ ensemble, where $N$ counts the number of molecules, but not the number of DOFs (which in our case are not strictly but only in average proportional). In the current implementation we use a Langevin thermostat, which acts without any interpolation on the atoms. The friction is proportional to the masses of the atoms and thus by themostatting each atom, the center of mass of every molecule obeys the fluctuation-dissipation-theorem as well ~\cite{adresso}.
A pure center of mass thermostat would also be sufficient, but slightly more difficult to implement in the integrator. The consequences for the definition of kinetic energy etc. that follow from "partial" degrees of freedom have been extensively discussed in \cite{jpa}.\\
The above discussion considers molecules which are replaced by their centers of mass in the coarse-grained region. However a more general mapping from explicit to coarse-grained, like a atomistic polymer mapped to a chain of spheres (so called "coarse-grained" beads) connected by springs, is possible. Then the interpolation is applied to the coarse-grained beads.

\section{Pressure calculation}
We can numerically analyze the local pressure in the stationary state and its behavior when an external field is applied  making use of the method proposed by Todd, Evans and Daivis \cite{todd1995pressure}. The time averaged local pressure tensor is given by
\begin{eqnarray}
{\bar p}_{\mu\nu}(x)=\frac{1}{2A \Delta x}\bigg\langle \sum_{\{ \alpha:|{{\bf X}_{\alpha}}-x|<\Delta x\}} ({\bf v}_{\alpha} \cdot \hat{{\bf e}}_\mu)({\bf v}_{\alpha} \cdot  \hat{{\bf e}}_\nu)  M_\alpha\bigg\rangle_t \\ +\frac{1}{2A}\bigg\langle \sum_{\alpha}{\bf F}_{\alpha}{\bf \hat{e}}_\mu \text{sgn}({\bf X}_\alpha\cdot {\bf \hat{e}}_x-x)\bigg\rangle_t
\label{prestensor}
\end{eqnarray}
where $A$ is a cross-sectional area perpendicular to the \emph{x}-direction. The first term corresponds to the ideal gas contribution calculated in a small slab centered at $x$ with thickness $2\Delta x$, and the second represents the contribution of the forces. Both sums run over molecules (which are labeled with $\alpha$ as before), where $M_\alpha$, ${\bf v}_\alpha$ and ${\bf X}_\alpha$ are mass, velocity and position. Consequently, the ideal gas contribution of the center of mass is fully included in the sum regardless of the weighting function of the molecule, since those degrees of freedom fully contribute to the average in any representation. The internal degrees of freedom, however, are not considered here. Therefore, the resolution of each molecule is implicit in the force contribution that depends on the weighting function. In the case of two-body forces, the second sum is reduced to the count of the forces that cross the area $A$. Particles far away from this plane (at distances bigger than the maximal cutoff of the non-bonded interactions) do not contribute to the sum.
\section{Simulation protocol}
The simulated system consisted of 8507 water molecules in a box of dimensions $16\;\text{nm}\times4\;\text{nm}\times4\;\text{nm}$ at ambient condition, the simulation was performed with a  modified GROMACS package \cite{hess2008gromacs}. Electrostatic interactions were treated with the reaction-field \cite{tironi1995generalized} approach using a cut-off length of 1.2\;nm. At a temperature of 300K and a pressure of 1\;bar we observe for a system
of 8507 water molecules a compressibility of
$\kappa_T^{at}=4.56\times10^{-10}m^2/N$\cite{wang2009comparative}.

The coarse-grained potential was obtained by mapping each water molecule in the center of mass to a single coarse-grained bead by an iterative Boltzmann inversion based on radial distribution functions. 100 steps of 0.2ns each were carried using the VOTCA software package \cite{ruhle2009versatile}. As mentioned before, such a coarse-graining does not preserve the state point, yielding a coarse-grained pressure of approximately 6200\;bar, while the compressibility is not altered. That is the local slope of the equation of state is conserved by such a procedure, while the absolute value is shifted. Additional corrections to reproduce the atomistic pressure can be applied \cite{reith2003deriving}, but have been shown to alter the compressibility \cite{wang2009comparative} by a factor of 5, which one has to avoid in order to obtain the correct particle number fluctuation in the different regions of space. To obtain the highest accuracy in the final density profile the thermodynamic force was extended by half the cut-off length (0.6\;nm) into the atomistic zone (because of the mixed atomistic coarse grained interaction of two molecules, when one molecule is in the transition region). Without this extension the density deviated at the most by about 1 \% from the target density, showing a drop in the atomistic zone. This is due to the fact that close to the coarse-grained zone, the atomistic molecules interact with the hybrid molecules an thus have a slightly altered state point.
\section{Diffusion profile}
\begin{figure}
\begin{center}
\includegraphics[width=0.5\textwidth]{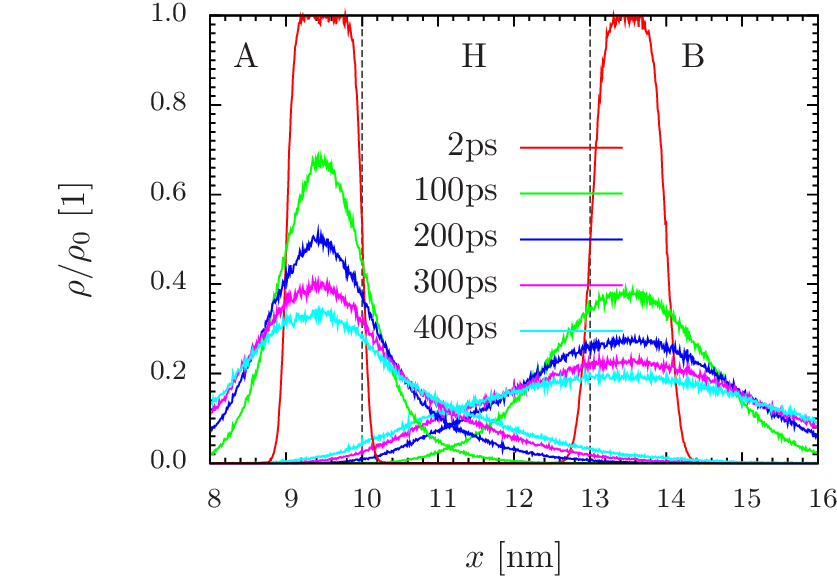}
\caption{
Interdiffusion between the two regions. This shows the density profiles over time for particles originally in a cubic subvolume with width $\Delta x=1\text{nm}$ at $t=t_0$. The profiles are asymmetric as diffusion constants are different in
 each representation. By this it is clear that both domains constantly exchange particles.}
\label{water_diffusion}
\end{center}
\end{figure}

To proof that molecules indeed diffuse back and forth between both regions without regard to the local level of resolution, we can measure diffusion profiles along the axis of resolution change. At $t=t_0$ all particles in a slab of width $\Delta x=1\text{nm}$ are selected which are then "traced" (in terms of their density profile) over time (fig. \ref{water_diffusion}). This is done for a set of particles in both regions. As shown by the symmetry of the decay for both cases, the transition region does not introduce any further transition barrier. On the coarse-grained side the particle distribution decays significantly faster, illustrating the enhanced diffusion constant in this region.
